# Quantitative phase-contrast imaging: a bridge between qualitative phase-contrast and phase retrieval algorithms

NATHANIEL HAI,* AND JOSEPH ROSEN

School of Electrical and Computer Engineering, Ben-Gurion University of the Negev, P.O Box 653, Beer-Sheva 8410501, Israel
*Corresponding author: mirilasn@post.bgu.ac.il

**In the last five decades, iterative phase retrieval methods draw large amount of interest across the research community as a non-interferometric approach to recover quantitative phase distributions from one (or more) intensity measurement. However, in cases where a unique solution does exist, these methods often require oversampling and high computational resources, which limits the use of this approach in important applications. On the other hand, phase contrast methods are based on a single camera exposure but provides only a qualitative description of the phase, thus are not useful for applications in which the quantitative phase description is needed. In this study we adopt a combined approach of the two above-mentioned methods to overcome their respective drawbacks. We show that a modified phase retrieval algorithm easily converges to the correct solution by initializing the algorithm with a phase-induced intensity measurement, namely with a phase contrast image of the examined object. Accurate quantitative phase measurements for both binary and continuously varying phase objects are demonstrated to support the suggested system as a single-shot quantitative phase contrast microscope.   © 2020 Optical Society of America**

From an overall perspective, quantitative phase imaging (QPI) techniques can be sorted into two main categories – interferometric and non-interferometric. The former approach is known for many years as the ad-hoc method for extracting the phase of electromagnetic waves [1], whereas the latter is gaining massive interest in the last five decades mainly due to the emerging of phase retrieval algorithms [2–4]. These algorithms aim to decipher the phase map of the object based on one intensity measurement on one plane, and a-priori knowledge on the intensity (or the complete intensity distribution) on the other plane, where the two planes are related via a known propagation operator [5]. This task is not trivial since the solution uniqueness is not guaranteed in all cases, and even when it does, oversampling of the signal is needed [5]. The ill-posed inverse problem of recovering phase information based on intensity measurements is partially solved by the transport of intensity equation (TIE) approach [6–8]. However, TIE requires at least three intensity distributions from neighboring axial planes, which decreases the acquisition rate of the method and increases its experimental complexity. The multiple camera exposures can be avoided by encoding the three-dimensional information of the object in a single-shot hologram [9], but this method involves interference process which complicates the optical system and classifies the technique in the old interferometric QPI. Although phase contrast methods [10–11] overcome these above-mentioned drawbacks and employ a single-shot on-axis scheme, they lack the numerical values of the imaged phase and cannot be used when QPI is required.

This Letter aspires to close the gap between the qualitative-only and the quantitative non-interferometric phase imaging techniques, by the prospect of a quantitative phase contrast microscope based on a two-step simple concept. First, a qualitative phase image of a pure phase object is acquired by one of the widely used phase contrast methods [10–11]. Then, the phase contrast image is fed into a digital framework based on a modified Gerchberg-Saxton algorithm (GSA) [2], which reproduces the optical system used to acquire the phase contrast image. The translation of phase information to intensity distribution at the image plane significantly increases the effectiveness of the phase retrieval algorithm, which is found to converge faster than conventional algorithms to the correct phase distribution of the object placed at the input plane. Although this is not the first time that iterative phase retrieval method is combined with phase-induced intensity measurements [12,13], the proposed quantitative phase contrast microscope is shown to be more accurate than the traditional interferometric approach in quantitative phase estimation [14]. Moreover, the experimental system is cost-effective and does not require sophisticated optical elements. It relies on standard refractive lenses, a phase shift plate, and a digital camera in a spatial filtering configuration that extremely simplifies the image acquisition process and provides inherent optical modularity. Another important benefit of the suggested method is the single-shot imaging which maximizes the temporal resolution of the method, and therefore can be used in phase imaging of dynamic scenes [15].

In the following, quantitative phase image reconstruction based on the two steps (Fig. 1) is described in more detail. Phase contrast image is formed by the telescopic optical setting illustrated in Fig. 1(a), which consists of two refractive lenses, a phase plate and an image sensor. The phase plate is designed to induce a phase shift of $\xi$ radians at the origin of the spectrum coordinates, which is related to the DC component of the phase object, so that the intensity emerging at the image sensor plane is given by,

$$
\begin{aligned}
I(\mathbf{r}) &= \left| \mathfrak{F}^{-1}\left\{ \mathfrak{F}\left\{ \exp\left[j\varphi\left(\frac{\mathbf{r}}{M_T}\right)\right] \right\} \cdot \left[1 - \delta(\mathbf{\rho}) + \delta(\mathbf{\rho})\exp(j\xi)\right] \right\} \right|^2 \\
&= \left| \mathfrak{F}^{-1}\left\{ \mathfrak{F}\left\{ 1 + \sum_{n=1} a_n \varphi^n\left(\frac{\mathbf{r}}{M_T}\right) \right\} \cdot \left[1 - \delta(\mathbf{\rho}) + \delta(\mathbf{\rho})\exp(j\xi)\right] \right\} \right|^2 \\
&\cong \left| \mathfrak{F}^{-1}\left\{ \delta(\mathbf{\rho})\exp(j\xi) + \mathfrak{F}\left\{ \sum_{n=1} a_n \varphi^n\left(\frac{\mathbf{r}}{M_T}\right) \right\} \right\} \right|^2 \\
&= \left| \exp(j\xi) + \sum_{n=1} a_n \varphi^n\left(\frac{\mathbf{r}}{M_T}\right) \right|^2 \\
&= 1 + \exp(j\xi)\sum_{n=1} a_n^* \varphi^n\left(\frac{\mathbf{r}}{M_T}\right) + \exp(-j\xi)\sum_{n=1} a_n \varphi^n\left(\frac{\mathbf{r}}{M_T}\right) \\
&\quad + \left| \sum_{n=1} a_n \varphi^n\left(\frac{\mathbf{r}}{M_T}\right) \right|^2,
\end{aligned}
$$

(1)

where $\varphi(\mathbf{r})$ is the phase of the object at the input plane with $\mathbf{r}$ as the transverse coordinates, $\delta(\mathbf{\rho})$ is the Kronecker delta-function with $\mathbf{\rho}$ as the transverse coordinates of the spectrum, $M_T$ is the transverse magnification of the system, $a_n$ is the $n$-th complex coefficient of the Maclaurin series expansion of $\exp[j\varphi(\mathbf{r})]$ and $\mathfrak{F}$ denotes the two-dimensional (2D) Fourier transform (FT). Under the small phase approximation [$\varphi(\mathbf{r})\ll1$ for all $\mathbf{r}$] the intensity of Eq. (1) becomes $I(\mathbf{r})\simeq1+2(\sin\xi)\varphi(\mathbf{r})$, but the proposed method is not limited only to a small phase distribution. Eq. (1) indicates that the second term of $I(\mathbf{r})$, which is the first and the larger term in the series is proportional to $\varphi(\mathbf{r})$. This observation is unlike the case of usual phase retrieval algorithms [5], where the measured signal is $I(\mathbf{r})=|\mathfrak{F}\{\exp[j\varphi(\mathbf{r})]\}|^2$, and experimental tests indicate that the information on $\varphi(\mathbf{r})$ is represented more by the phase of $\mathfrak{F}\{\exp[j\varphi(\mathbf{r})]\}$, rather than by its magnitude [5,16]. From Eq. (1) we conclude that the difference between the background phase and the object phase arriving the image sensor plane can be tuned by changing $\xi$ to increase the contrast of the phase object located at the object plane. Once the phase contrast is maximized, the captured intensity image undergoes multiple digital propagations between the image and object planes in the modified GSA illustrated in Fig. 1(b) as follows. The algorithm is initialized with a complex amplitude of constant phase values and a magnitude equals to the square root of the recorded phase contrast image. This complex amplitude is then propagated to the object plane by a sequence of 2D FT followed by division with the phase plate followed by 2D inverse FT. The obtained phase in the object plane is propagated to the image plane by using a similar sequence. This iterative process is continued until convergence of the phase at the object plane is accomplished. In other words, the recovered phase at the object plane is iteratively propagated between the image and object planes as long as the condition,

$$\frac{1}{S}\int\left[\hat{\varphi}_n(\mathbf{r})-\hat{\varphi}_{n-1}(\mathbf{r})\right]^2 d\mathbf{r} > \Delta, \quad (2)$$

is satisfied. $\hat{\varphi}_n(\mathbf{r})$ is the recovered object phase in the $n$ iteration of the modified GSA run, $S$ is the area of the phase object and $\Delta$ is a pre-determined threshold. One can easily validate that the modified GSA employed in the digital reconstruction, is the exact description of the optical system used to acquire the phase contrast image. It is important to note that although matrix division is not a recommended computational operation that might give rise to singular values, here the division is by a pure phase matrix which is equivalent to multiplication with a matrix of phase conjugated of the original phase plate. Hence, singular values are impossible to appear during the quantitative phase reconstruction in the computer.

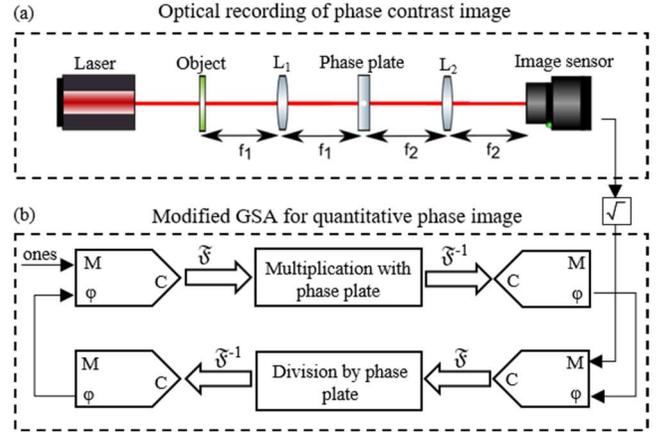

Fig. 1. Workflow of the suggested quantitative phase contrast microscope. (a) Optical configuration used to record a phase contrast image of pure phase objects, which is then fed to (b) a modified GSA, based on the optical system, to reconstruct the quantitative phase values at the object plane. In the current implementation L$_1$ is a microscope objective.

We experimentally tested the proposed method capability in QPI task of two kinds of phase objects, one exhibits a sharp change between two phase values and the second contains a gradual variation in phase values. The examined object was illuminated by a HeNe laser (AEROTECH, maximum output power of 25 mW @ $\lambda$ = 632.8 nm) and transformed to the spectrum domain by a microscope objective (MO, Olympus PLN 10X, NA = 0.25). In the spectrum domain, the signal was multiplied by the phase plate implemented by a spatial light modulator (SLM, Holoeye PLUTO-2, 1920×1080 pixels, 8 μm pixel pitch, phase only modulation). The SLM displayed a phase pinhole of $\xi$ radians, realized in its 6×6 central pixels with phase value of $\xi$ surrounded by constant phase of zeros for the entire pixels. The size of the phase pinhole was chosen according to the Rayleigh criterion, so that the phase pinhole modulation will only apply on the zero order of the spectrum and the higher orders phase will remain unaltered. A refractive lens ($f$=500 mm, $D$=50.8 mm) transformed the spectrum signal multiplied by the phase plate back to the spatial domain, and the phase contrast image was captured by the image sensor (Thorlabs 8051-M-USB, 3296×2472 pixels, 5.5 μm pixel pitch) according to Eq. (1). Note that the output reliability of the digital step in the method, that is the reconstructed quantitative phase image, is highly dependent on the phase contrast image acquired in the optical step. Therefore, the recorded phase contrast image must be a decent qualitative equivalent of the phase object under examination. To achieve this goal, the phase shift value $\xi$ was adjusted to obtain a correct phase contrast image, and a value of $\xi=\pi/2$ was inspected to derive the best results. Moreover, the laser beam was filtered and collimated, in order to approach as close as possible to a plane wave illumination of the object. This prevents from unwanted aberrations to interfere the recorded phase contrast image.

For the first demonstration, a USAF 1951 phase target (Quantitative Phase Microscopy Target, Benchmark Technologies) was placed at the object plane. The target elements and the substrate on which they are deposited have their refractive index matched at 1.52, so that the phase delay between them is solely induced by the elements thickness. Phase contrast images that were acquired by the optical system for element thicknesses of 150 nm and 250 nm, are shown in Figs. 2(c) and 2(d), respectively. For comparison, Fig. 2(a) illustrates the intensity image without filtering with the phase plate. As expected, the intensity image of Fig. 2(a) has no phase contrast and therefore carry no information regarding the phase of the examined transparent object. Next, we have to convert the phase-induced intensity measurements to a reliable quantitative phase information. This can be done by initiating the modified GSA, unique to the used optical system, with the phase contrast measurement and iteratively solve for the phase at the object plane. Since the captured image size and the phase plate size (realized by the SLM) are different, a central region of interest from the recorded phase contrast image in the size of the phase plate, was cropped and used as the modified GSA input. The fact that the phase information at the object plane is presented in the phase contrast image, which is fed to the iterative algorithm, has two important consequences. First, the iterative process is found to converge only after 6 iterations to the correct answer with a threshold value of less than 1 milliradian$^2$, according to Eq. (2) (see Fig. 3). Second, the reconstructed phase is more accurate compared to the retrieved phase when the algorithm is initialized with a non-filtered image [Fig. 2(a)] or low phase contrast image ($\xi \neq \pi/2$). Figs. 2(e) and 2(f) illustrates the reconstructed phase, which is the output of the modified GSA for element thickness of 150 nm and 250 nm, respectively. As an immediate quality check, one can verify the thickness differences for the two objects by comparing the contrast between the features and the background, which is higher for the 250 nm thickness object as expected. Nevertheless, we are interested in assessing the quantitative certainty of the method rather than the qualitative validity. Therefore, the element thickness was estimated by the corresponding phase delay from the reconstructed phase map [Figs. 2(e) and 2(f)]. To achieve the best estimation, the element phase was calculated on the largest feature in the phase target (yellow square) and the background phase was calculated on the largest flat background region (green rectangle). The mean phase value of these two regions for the 150 nm thickness target is 2.68 and 1.84 radians for the yellow square and green rectangle, respectively. The mean phase value of these two regions for the 250 nm thickness target is 2.97 and 1.71 radians for the yellow square and green rectangle, respectively. Subtraction of these values for each target accounts for the phase delay of the element, from which the element thickness can be estimated by dividing it by the product between the illumination wavenumber and the difference between the object and air refractive indices. Estimation of the 150 nm and 250 nm objects thickness based on the reconstructed phase map, was calculated to be 161 nm and 244 nm, respectively. These estimations are closer to the reported manufacturer's values than the estimations achieved by an interferometric approach for the same phase objects (see Ref. [14]).

In order to further verify the reliability of the suggested framework as an effective QPI tool, the phase contrast image of Polystyrene microspheres (Focal Check, 6 μm diameter) was captured by the optical system and the quantitative phase map was reconstructed by the modified GSA. The specimen is constructed so that the microspheres are having a refractive index of 1.587 and sealed within a medium of a 1.56 refractive index, so that the expected maximal phase delay of 1.61 radians is obtained at the structure center. The reconstructed phase of a region from the sample that contains two microspheres is shown in the 3D surface plot in Fig. 2(b). The inset of Fig. 2(b) shows the corresponding 2D upper view, where the phase delay associated with each of the spherical features shows good symmetry and smoothly varying phase delay as expected from a gradual change in the thickness of the phase object. To examine the reconstructed phase values of the microspheres, estimation of each spherical structure diameter was carried out based on the phase delay values. Similar to the previous demonstration, the background phase was calculated on the reconstructed phase portion that does not contain the microspheres and found to equal 1.95 radians. This value was subtracted from the phase value at the center of each sphere, 3.57 radians for both. From the result in phase delays, each structure diameter was calculated and found to be 6.04 μm. The good agreement between the phase values estimated using the suggested quantitative phase contrast microscope and the reported phase values, in two independently instances of different types of phase objects is summarized in Table 1. The nature of convergence of the recovered phase from the modified GSA for the three experimentally measured objects and a simulated object is described in Fig. 3. It should be emphasized that all of the phase objects used in our experiments have phase variations of less than $2\pi$ radians, and therefore phase-unwrapping procedure is not necessary [17]. In case that a phase object of optical thickness (physical thickness multiplied by the refractive index) larger than the illumination wavelength is under examination, the output of the modified GSA will be given to $2\pi$ ambiguities due to the cyclic nature of the trigonometric functions. These ambiguities can be removed by the use of a suitable phase-unwrapping algorithm.

In summary, we constructed a new QPI method that consists of two stages. One is the recording of optical signal and the other is iterative phase reconstruction based on the recorded phase contrast image. The main contribution of this study is the bridge between phase retrieval approach which in many cases lacks prior information on the examined object that hardens the phase reconstruction task [5], and the phase contrast approach which lacks the quantitative phase information. The proposed method takes the advantages of each of the two aforementioned approaches to compensate for the disadvantages of the other, and brings them together to generate an accurate and robust QPI apparatus. We show that by using the suggested quantitative phase contrast microscope, accurate reconstruction of phase objects can be achieved from a single camera exposure and within only few iterations of the modified phase retrieval algorithm. The relatively simple optical configuration and low computational demands, in addition to the single camera shot recording and the accurate quantitative phase estimation qualify the suggested technique for label-free imaging of transparent micro-organisms and inspection of non-absorbing elements.

**Table 1. Thickness estimation of phase objects based on the quantitative phase reconstructions using the suggested approach**

| Phase object | Estimated thickness |
|---|---|
| USAF 1951 phase target (150 nm) | 161 nm |
| USAF 1951 phase target (250 nm) | 244 nm |
| Polystyrene microspheres (6 μm diameter) | 6.04 μm |

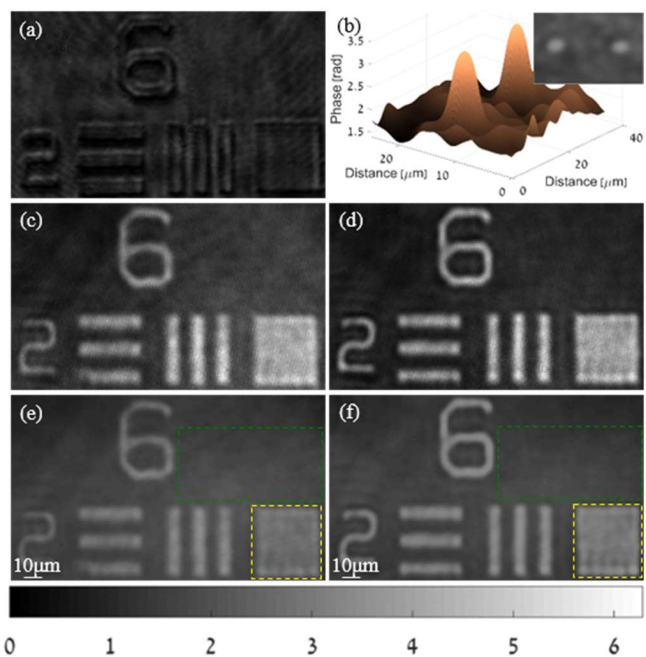

Fig. 2. Experimental demonstration of the suggested quantitative phase contrast method. (a) Intensity image captured by the camera without the phase contrast operation and (c), (d) phase-induced intensity images of USAF 1951 phase targets captured by the camera with the phase contrast operation, for target thickness of (c) 150 nm, and (d) 250 nm. (e), (f) Quantitative phase image obtained from the modified GSA output after initialization with the corresponding phase contrast images (c), (d). The green rectangle and yellow square correspond to the regions where the background and element phases were calculated, respectively. (b) 3D surface phase plot of the 6 μm diameter Polystyrene microspheres reconstructed in a similar method to the USAF 1951 phase target (phase contrast image is not shown). Upper 2D view is shown in the inset. Horizontal color bar corresponds to phase values of images (e) and (f) and the inset of (b) in radians.

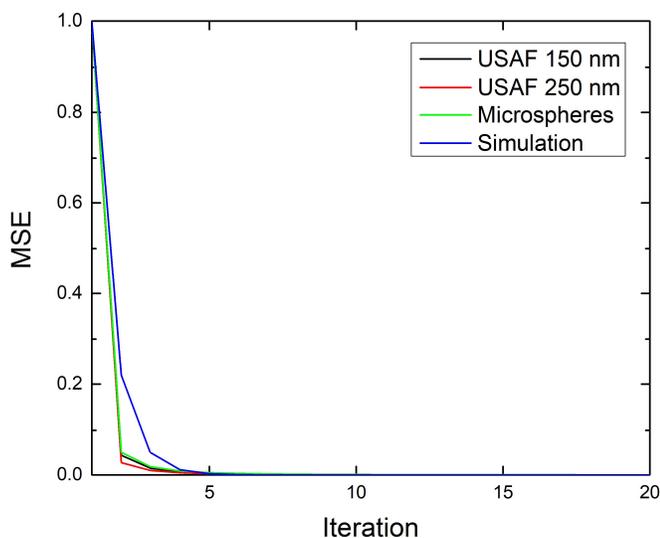

Fig. 3. Convergence plots of the recovered phase distribution from the modified GSA. Normalized mean square error (MSE) as a function of the iteration number for the three examined objects and a simulated object shows that convergence is achieved within few iterations. MSE for the experimental measurements was calculated according to Eq. (2), while for the simulated object $\hat{\varphi}_{n-1}(r)$ from Eq. (2) was replaced with the actual phase of the object.

**Funding**. The ATTRACT project funded by the EC under Grant Agreement 777222; Israel Science Foundation (ISF) (1669/16); Ministry of Science, Technology and Space.

**Acknowledgment.** The authors thank Dr. A. Vijayakumar for fruitful discussions in the beginning of this research.

**Disclosures.** The authors declare no conflicts of interest.